\documentclass[
    journal,
    twocolumn,
    ]{IEEEtran}
\usepackage{graphicx}
\usepackage{amsmath}
\usepackage{amssymb}
\usepackage{epsfig}
\usepackage{epstopdf}
\usepackage{amsthm}
\usepackage{afterpage}
\usepackage{verbatim}
\usepackage{psfrag}
\usepackage{color}
\usepackage{cite}
\usepackage{setspace}
\usepackage{adjustbox}
\usepackage{epsfig}
\usepackage{epstopdf}
\usepackage{footmisc}
\usepackage{fmtcount}
\usepackage{stackrel}
\usepackage{float}
\usepackage{breqn}
\usepackage{hyperref}
\usepackage{enumerate}
\usepackage{graphicx}
\usepackage{subcaption}
\usepackage{graphicx}

\usepackage{tabularx}

\begin{document}

\title{Point-to-Point Communication in Integrated Satellite-Aerial Networks: State-of-the-art\\ and Future Challenges}
\author{Nasir Saeed, ~\IEEEmembership{Senior Member,~IEEE,} Heba Almorad, ~\IEEEmembership{Student Member,~IEEE,} Hayssam Dahrouj, ~\IEEEmembership{Senior Member,~IEEE,} Tareq Y. Al-Naffouri, ~\IEEEmembership{Senior Member,~IEEE,} Jeff~S.~Shamma, ~\IEEEmembership{Fellow,~IEEE,} and Mohamed-Slim Alouini, ~\IEEEmembership{Fellow,~IEEE}
\thanks{Nasir Saeed, Tareq Y. Al-Naffouri, Jeff S. Shamma and Mohamed-Slim Alouini are with the Department of Computer, Electrical and Mathematical Sciences and Engineering (CEMSE), King Abdullah University of Science and Technology (KAUST), Thuwal, Makkah Province, Kingdom of Saudi Arabia, 23955-6900. \newline
	Heba Almorad is with the Department of Electrical and Computer Engineering, Effat University, Jeddah 22332, Saudi Arabia. \newline
    Hayssam Dahrouj and Jeff S. Shamma are with the Center of Excellence for NEOM Research, King Abdullah University of Science and Technology, Thuwal 23955-6900, Saudi Arabia.	
}
}

\maketitle

\begin{abstract}
This paper overviews point-to-point (P2P) links for integrated satellite-aerial networks, which are envisioned to be among the key enablers of the sixth-generation (6G) of wireless networks vision. The paper first outlines the unique characteristics of such integrated large-scale complex networks, often denoted by spatial networks, and focuses on two particular space-air infrastructures, namely, satellites networks and high-altitude platforms (HAPs). The paper then classifies the connecting P2P communications links as satellite-to-satellite links at the same layer (SSLL), satellite-to-satellite links at different layers (SSLD), and HAP-to-HAP links (HHL). The paper overviews each layer of such spatial networks separately, and highlights the possible natures of the connecting links (i.e., radio-frequency or free-space optics) with a dedicated overview to the existing link-budget results. The paper, afterwards, presents the prospective merit of realizing such an integrated satellite-HAP network towards providing broadband services in under-served and remote areas. Finally, the paper sheds light on several future research directions in the context of spatial networks, namely large-scale network optimization, intelligent offloading, smart platforms, energy efficiency, multiple access schemes, and distributed spatial networks.
\end{abstract}

\section{Introduction}

Connectivity is the backbone of modern digital economy with over three billion people connected worldwide, and more than 14 billion devices connected through the Internet core network. Although the wireless coverage has spread substantially over the past two decades, almost half of the world's population remains unconnected \cite{hernandez}. With the data deluge in terms of global services and user-equipments, the number of connected devices is expected to surpass 50 billions, which poses stringent burdens on the current telecommunications terrestrial infrastructure \cite{hernandez}. Therefore, developing novel connectivity solutions to fulfill such enormous demands becomes an indispensable necessity.

A recent trend for boosting ground-level communication is by enabling connectivity from the sky as a means to connect the unconnected and super-connect the already connected, a theme that falls at the intersection of the ongoing sixth-generation (6G) wireless networks initiatives \cite{dang2020should, yaacoub2020key, saeed2019cubesat}. Towards this direction, integrated satellite-aerial networks, also known as spatial networks (SNs), have emerged as essential enablers for serving remote areas and enhancing the capacity of the existing wireless systems \cite{dang2020should,yaacoub2020key, saarnisaari20206g, 8760401, saeed2019cubesat}. Thanks to their capabilities at connecting wireless platforms of different altitudes, SNs provide high data rates for terrestrial wireless backhaul networks \cite{abu2019performance}, and enable global Internet services \cite{qiu2019air}. While the original focus of SNs is mainly on satellites deployment, recent SNs studies include other non-terrestrial networks that operate at a comparatively lower altitude, i.e., communications infrastructures at the stratosphere and troposphere layers \cite{Rinaldi2020}. Besides connectivity, SNs have plenty of valuable applications, e.g., surveillance, weather forecasting, earth observation, navigation, and climate monitoring \cite{czichy2000inter, saeed2020wireless, Saeed2020IoT}.

\begin{figure*}[h]
    \centering
    \includegraphics[width=\linewidth]{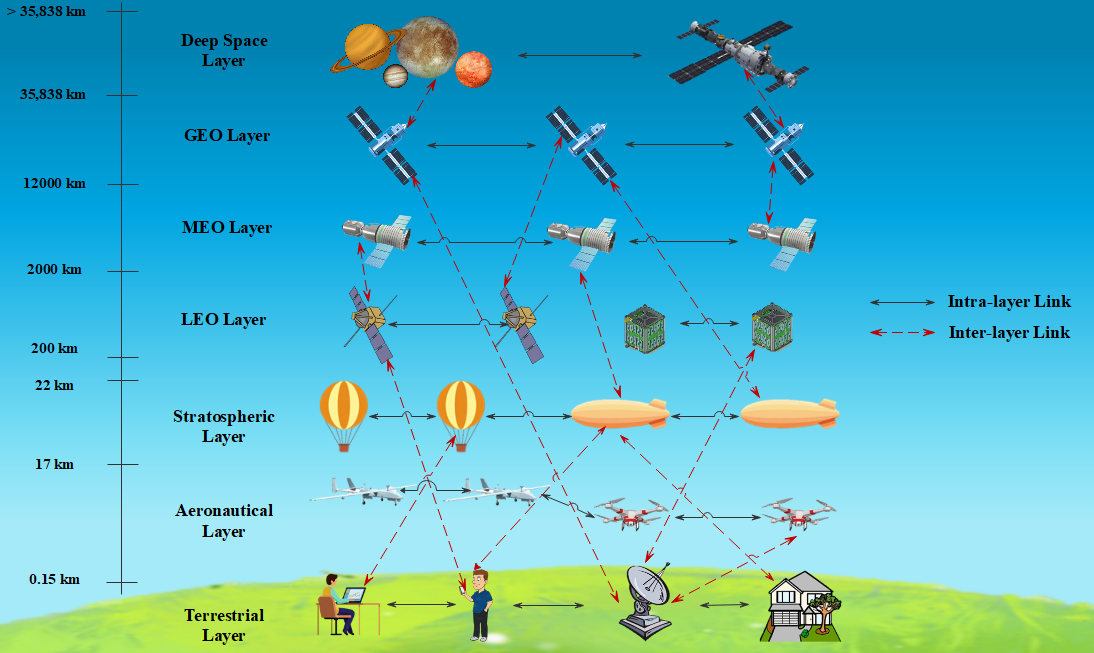}
    \caption{Illustration of a multilayered SN with satellites, HAPs, and UAVs.}
     \label{multi}
\end{figure*}

Spatial networks consist of a plurality of nodes (also called spatial elements) in two- and three-dimensional spaces, which form single and multilayer architectures. Such nodes can be satellites, high-altitude platforms (HAPs), tethered balloons, or unmanned aerial vehicles (UAVs) \cite{barthelemy_2014}. The type of architecture then depends on the altitude of nodes. While the nodes at the same altitude are called single-layer nodes, the nodes at different altitudes are called multilayer nodes. The multilayered architecture often offers more degrees of freedom than the single-layer, and can provide a global connectivity solution since the multilayered architecture combines several layers, and exploits the compound benefits of the different layers at the different altitudes \cite{akyildiz2002mlsr}. Fig. \ref{multi} illustrates a generic multilayered architecture of SNs where each layer is at a different altitude from the Earth's surface, i.e., deep space ($>$ 35,838 km), geo-synchronous Earth orbit (GEO) (12000-35,838 km), medium Earth orbit (MEO) (2000-12000 km), low Earth orbit (LEO) (200-2000 km), stratospheric (17-22 km), and aeronautical (0.15-17 km) \cite{cao2018airborne}.  The spatial elements in each layer can relay data in a multihop fashion among the different nodes of SNs, thus converting a long-range single-hop link into short-range multi-hop links, thereby reducing the overall propagation delay and improving the overall data rate \cite{DETTMANN2018187}.

The multi-hop links can be established within a single layer (intra-layer) of SNs or between nodes of two or more different layers (inter-layer), as illustrated in Fig. \ref{multi}.  One can then categorize the SNs communications links as satellite-to-satellite links at the same layer (SSLL), satellite-to-satellite links at different layers (SSLD), HAP-to-HAP links (HHL), and UAV-to-UAV links (UUL), respectively.  Satellites, HAPs, and LAPs are equipped with on-board processing (OBP) capabilities to establish such links,  allowing the communication between different elements on the same layer or even at different layers in SNs  \cite{ares}. One significant difference between the terrestrial networks and SNs is that the latter consists of network topologies with significantly heterogeneous network nodes within the well-spread space-air layers, as illustrated in Fig. \ref{multi}. The links in such a multilayer network can be established using both radio-frequency (RF) waves and free-space optics (FSO), as discussed in details later in the paper.

In the current practice, radio frequencies in the microwave band are used to establish point-to-point (P2P) wireless links among the different entities of SNs. For example, the common data link (CDL) that is designed by the U.S Department of Defense uses Ku (12-18 GHz) and Ka (26-40 GHz) frequency bands to transmit data for long P2P communication between HAPs and terrestrial stations  \cite{yang2007next}. However, CDL's limited spectrum constraints limit its data rate between 274 Mbps to 3 Gbps, which do not satisfy the demand for high-speed wireless links  \cite{8760401} \cite{yang2007next}. In this context,  U.S. Defense Advanced Research Projects Agency (DARPA) started a program called ``Free-space Optical Experimental Network Experiment (FOENEX)" to develop links that can transmit data using FSO at a much higher speed. In 2012,  FOENEX successfully established the first FSO link to allow a 10 Gbps transmission rate for airborne platforms. After further improvement, it turned out that FSO can provide up to 100 Gbps P2P links using wavelength-division multiplexing (WDM), which is superior than the average rates of RF-based systems \cite{juarez2012analysis}. FSO technology is also energy-efficient, secure, and license-free, which make it a strong candidate for space-borne P2P communication deployment \cite{kaushal2016optical} \cite{williams2007rf}. FSO technology is, however, generally vulnerable to the environment and cannot operate efficiently in a rainy, snowy, or foggy weather. Also, the FSO links require perfect alignment between the transmitter and receiver of the moving platforms \cite{henniger2010introduction}, which is often handled using a variety of alternative techniques \cite{Douik2016, gibalina2017optical, horkin1998future}.
Consequently, DARPA launched another program to investigate ways of establishing the same 100 Gbps with all-weather tolerance capability. Towards this direction, the program investigated the mmWave spectrum (30-300 GHz) and exploited high-order modulation and spatial multiplexing techniques to attain the desired data rate for a range of 200 km intra-layer link, and 100 km for the inter-layer link in the stratospheric region \cite{agency_information_2020}. DARPA then identified mmWave technology as the suitable solution for airborne communication. The results showed an outstanding performance achieving 100 Gbps under the atmospheric attenuation, and cumulus loss with less than 0.3 dB/km in the E-band (71–76 GHz and 81–86-GHz).

Other interesting ongoing projects on SNs P2P links adopt hybrid RF/FSO \cite{malinowski2010high}, as a means to combine the mutual advantages of both RF and FSO. Such systems operate by switching to low-capacity RF links in bad weather conditions, or to high-capacity  FSO links under perfect transceivers alignment and suitable weather conditions. One such hybrid project is Integrated Aerial Communications (FaRIA-C) headed by DARPA \cite{fso}. This project started in 2019 to develop simultaneous hybrid links that switch between FSO and RF, based on the environment suitability. In other words, whenever the weather obscures the Line-of-Sight (LoS), the system switches from FSO to RF. FaRIA-C achieves up to 10 Gbps link capacity when operating at FSO and 2 Gbps at RF band  \cite{fso}. Despite their promising capabilities, hybrid FSO/RF systems still face various challenges, such as scheduling, scalability of the network, and quality of service (QoS) constraints, as highlighted in \cite{stotts2009hybrid}. In Table I, we summarize some of the well-known projects that use different communication technologies for enabling P2P links in SNs.

\begin{table*}[ht]
\caption{List of a few projects that uses P2P wireless communication links.}\label{table:proj}
\centering
\begin{tabular}{ |p{2.3cm}|p{2.5cm}|p{1.5cm}|p{3.5cm}|p{2.7cm}|p{1.4cm}|p{1cm}| }
 \hline
 \hline
\textbf{Project} & \textbf{Technology} & \textbf{Platform}& \textbf{Link type} & \textbf{Data rate} & \textbf{Distance (km)} & \textbf{Year} \\
\hline
\hline
Iridium \cite{inproceedings}      &RF (L-band) &Satellites &LEO-to-LEO   &25 Mbps   &  -    &1997    \\
\hline
SILEX \cite{silex}             & FSO (847nm-819nm) &Satellites  &GEO-LEO Link    & 50 Mbps  & 45000   &2001       \\
\hline
IRON-T2 \cite{stotts2009hybrid}     &  FSO (1556.1nm) / RF (X/Ku-band)    &- &LAP-to-LAP &2.5-40/0.274 Gbps   & 50-200 &2007       \\
\hline
 FALCON \cite{thompson_2020}      & FSO &Aircrafts & LAP-to-LAP &2.5 Gbps & 130&2010       \\
 \hline
LAC \cite{laser}         &  FSO (532 nm) &Airships, UAVs & HAP-to-HAP, LAP-to-LAP, and HAP-to-LAP & 10-40 Gbps & 200 & 2014   \\
\hline
CURfEGC \cite{meo}       &  RF (UHF) &Satellites & MEO-to-MEO &1-2 Mbps  &31,400 &2016 \\
\hline
QB50 \cite{qb50_2020}   & RF (VHF/UHF) &Satellites &LEO-to-LEO &0.5-10 kbps &90 &2017\\
\hline
Stellar \cite{Sansone2020}   & FSO (915nm) &Satellites &LEO-to-LEO &100 Mbps &1000 &2020\\
\hline
\hline
\end{tabular}
\end{table*}

\subsection{Related Review Articles}
Due to the significance of P2P communications in SNs, there is a plethora of review articles, each discussing different aspects of SNs \cite{1423332, 5464268,6248649,BEKMEZCI20131254,6963085,7317490,radhakrishnan2016survey,kaushal2016optical,son2017survey,8411465,8648737,cao2018airborne, 8675384,arum2020review,kodheli2020satellite,saeed2019cubesat}. For instance, reference \cite{BEKMEZCI20131254} reviews UAVs-based ad hoc networks, including the application scenarios, design characteristics and considerations, communication protocols, and open research issues. Chen et al. provide a  survey focusing on the coverage problem in UAV networks until 2014 \cite{6963085}. Then, reference \cite{7317490} further extends the literature on UAV communication and coverage issues such as routing, seamless handover, and energy efficiency until 2016. \cite{8675384} presents an updated UAV communications survey that discusses the practical aspects, standardization advancements, and security challenges. Furthermore, the authors in \cite{8675384} enumerate the 3GPP study items for maximizing the UAV opportunities in 4G and 5G applications. Moreover, \cite{8411465} surveys channel modeling for UAV communications, including channel characterization, channel modeling approaches, and future research directions.

From the stratospheric layer perspective, reference \cite{1423332} explores various facets of  P2P wireless communications in HAPs, including channel modeling, interference, antennas, and coding. The study in \cite{5464268}  is further narrowed down to FSO for P2P wireless links in HAPs, mainly focusing on acquisition, tracking, and pointing (ATP) issues. Recently, the authors in \cite{arum2020review} present a comprehensive and up-to-date survey on how to extend coverage and resolve capacity issues in rural areas using HAPs. The focus in \cite{arum2020review} is on HAPs regulations, projects, network topologies, and handover mechanisms. Moreover,  the authors in \cite{cao2018airborne} conduct extensive research on heterogeneous SNs, i.e., HAPs and LAPs, but does not come across the satellites aspects of SNs.

Reference \cite{kaushal2016optical} presents more detailed insights on SNs, such as ATP for space-based optical links, hybrid RF/FSO solution, MIMO, and adaptive optics. Unlike the above articles, the review \cite{kaushal2016optical} addresses all the layers of SNs; however, it focuses mainly on the satellites layer by discussing various satellite system aspects, medium access protocols, networking, testbeds, air interface, and future challenges \cite{kodheli2020satellite}.

In terms of space networks, Mukherjee et al.  survey the communication technologies and architectures for satellite networks and interplanetary Internet, demonstrating the notion of delay-tolerant networking (DTN) for deep space networks \cite{6248649}. Furthermore, Krishnan et al.  present an extensive study on diverse inter-satellite link design issues based on the last three layers of the open system interconnection (OSI)  \cite{radhakrishnan2016survey}.  \cite{radhakrishnan2016survey} proposes employing DTN protocols as a solution to the problems surveyed, detailing the required design parameters for inter-satellite communications. Moreover, dynamic resource allocation algorithms and schemes in integrated GEO satellite-ground networks are reviewed in \cite{8648737}.  \cite{saeed2019cubesat} highlights various issues for small satellites called CubeSats, discussing the coverage, different constellation designs, upper layer issues, and future research challenges. Moreover, \cite{kaushal2016optical} and \cite{son2017survey} present a study on FSO communications for satellites, including uplinks, downlinks, and ISL links. In Table \ref{table:lit}, we summarize the contributions of related review articles.
\begin{table*}
\caption{Comparison of this paper with the existing surveys.}\label{table:lit}
\centering
\begin{tabular}{|p{3cm}|p{1.4cm}|p{11.0cm}|p{0.7cm}|}
 \hline
 \hline
\textbf{Ref.}  & \textbf{Platform Type} & \textbf{Area of Focus} & \textbf{Year} \\
\hline
 \hline
Karapantazis et al.\cite{1423332} &HAPs & Presents possible architectures of HAPs, system structure, channel modeling, antennas, coding techniques, resource allocation techniques, and applications.  &2005 \\
\hline
Fidler et al. \cite{5464268}  &HAPs & Outlines FSO communication technology, system design requirements, data transmission and correction techniques, and experimental field trials for HAPs. &2010 \\
\hline
Mukherjee et al. \cite{6248649}   &Satellites& Discusses architectures, communication technologies, networking protocols, interplanetary Internet, and open research challenges for satellite networks. &2013 \\
\hline
Bekmezci et al. \cite{BEKMEZCI20131254}  &LAPs & Focuses on design characteristics, routing protocols, applications and open research issues for UAV networks. & 2013 \\
\hline
Chen et al. \cite{6963085}  &LAPs & Discusses the coverage issues for  UAV networks. &2014 \\
\hline

Gupta et al.\cite{7317490}  &LAPs & Reviews major issues in UAV communication networks, including mobility, limited energy, and networking. &2016 \\
\hline
Krishnan et al. \cite{radhakrishnan2016survey}   & Satellites & Presents various design parameters based on the last three OSI model for small satellite networks, such as modulation-and-coding, link design, antenna type, and different MAC protocols.  &2016 \\
\hline
Kaushal et al. \cite{kaushal2016optical}  &Satellites &Discusses space-based optical backhaul links and their applications. &2017\\
\hline
Son et al. \cite{son2017survey}   &Satellites &Highlights the importantence of FSO communications for inter-satellite links. &2017 \\
\hline
Khuwaja et al.\cite{8411465}  &LAPs & Surveys channel modeling techniques for UAV communications.&2018 \\
\hline
Peng et al. \cite{8648737}  & Satellites & Presents dynamic resource allocation schemes for integrated satellite and terrestrial networks. &2018 \\
\hline
Cao et al. \cite{cao2018airborne}  &HAPs/ LAPs & Discusses design parameters and protocols for HAPs and LAPs communication networks. &2018 \\
\hline
Fotouhi et al.\cite{8675384}  &LAPs & Surveys practical aspects, standardization, regulation, and security challenges for UAVs-based cellular communications. &2019 \\
\hline
Arum et al. \cite{arum2020review}  &HAPs & Focuses on coverage and capacity issues using HAPs. &2020 \\
\hline
Saeed et al. \cite{saeed2019cubesat}  &Satellites &  Presents channel modeling, modulation-and-coding, coverage, constellation issues, networking, and future research directions for CubeSat communications. &2020 \\
\hline
Kodheli et al. \cite{kodheli2020satellite}  &Satellites &Discusses the recent technical advances in scientific, industrial, and standardization for satellite communications. &2020 \\
\hline
This paper  & SNs & Outlines the unique characteristics of large-scale complex SNs, surveys various wireless communication technologies to implement P2P links' in SNs, and points out several promising research directions. &2020  \\
\hline
\hline

\end{tabular}
\end{table*}

\subsection{Contributions of our Paper}
Unlike the above-mentioned surveys which only focus on a single non-terrestrial network layer, i.e., either satellites or HAPs, our current paper focuses on P2P links for a multi-layered spatial network. The main motivation of this survey originates from their importance of studying the unique characteristics of spatial networks and the P2P interconnecting links in light of 6G large-scale complex networks. To this end, the paper presents the studies on wireless communication technologies for each layer separately, including satellites and HAPs layers. In conjunction, the paper overviews two possible alternatives for intra- and inter-satellite links, mainly FSO and RF connections, and discusses various possibilities for enabling P2P links among HAPs and from HAPs to the ground station. To best illustrate the compound benefits of the different layers integration, the paper then sheds light on the integrated satellite-HAP network as a means to provide broadband services in underserved areas. Finally, the paper presents several future research directions in the context of spatial networks, including large-scale network optimization, intelligent offloading, smart platforms, energy efficiency, multiple access schemes, and distributed spatial networks.
\subsection{Paper Organization}
The rest of the paper is organized as follows. Section II presents P2P links in satellite networks, covering both intra- and inter-layer links. Moreover, it provides link budget calculation for both RF and FSO-based inter-satellite links. We report the studies on P2P links in HAP-based networks in Section III, discussing both inter-HAP links and HAPs-to-ground communication. Section IV provides a review of integrated satellite-HAP networks to improve the reliability, coverage, and scalability for future 6G wireless communication systems. We present numerous future research directions in Section V, and then we conclude the paper in Section IV.

\begin{figure*}[h]
	\centering
	\includegraphics[width=\linewidth]{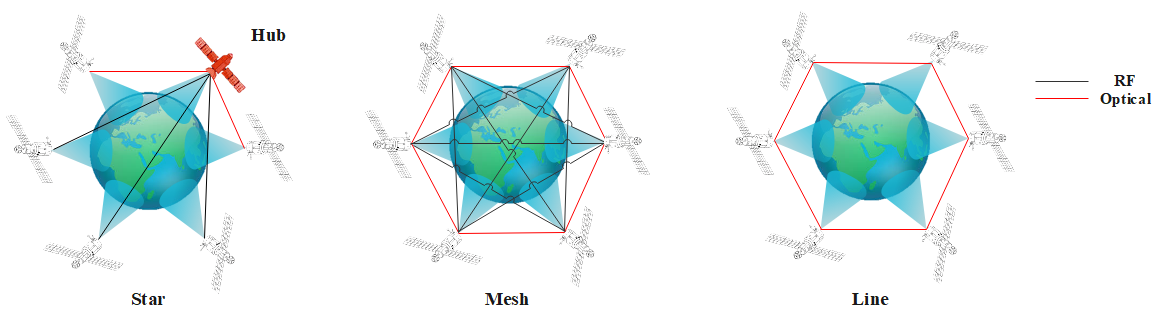}
	\caption{Illustration of different satellite topologies with satellite-to-satellite links at same layer.}
	\label{top}
\end{figure*}

\section{P2P Links in Satellite Networks}
With the emergence of the new space economy, satellite communication is getting more advanced in providing the Internet from space.  The satellite networks consist of many satellites at different altitudes, revolving in various types of constellations, using different frequency bands with distinct coverage. Therefore, it is critical for the satellite networks to take into account the essential characteristics, such as altitude, constellation, and operating frequency band, to achieve a specific goal. For example, the higher the satellite is, the wider the area it covers (a GEO satellite can cover around 30$\%$  of the Earth's surface, while a group of MEO and LEO satellites is required to cover the same area). On the other hand, MEO and LEO satellites provide shorter paths than GEO, resulting in less propagation delay. Also, satellites in low altitude constellations move faster, leading to a higher Doppler effect. Besides, the GEO, MEO, and LEO, constellations can be designed in such a way to increase the dwell time in certain parts of the world, for example, in highly elliptical orbits (HEO) \cite{kodheli2020satellite}.

Apart from the constellation design, enabling P2P links among the satellites is crucial for relaying the data. There are two possible relaying methods in satellite networks, namely amplify-and-forward (AF) and decode-and-forward (DF) \cite{Bhatnagar2015}. Satellites that use AF techniques are known as transparent satellites because they only amplify the received signal and forward it to the neighboring satellites or the ground station. On the other hand, DF satellites, or regenerative satellites, decode the incoming signal and perform signal processing to mitigate the interference and regenerate it.
Besides relaying, the selection of a routing topology is critical for efficient communication between the satellites and the ground segments, or between the satellites. Typically, there are three topologies (i.e., star, mesh, and line) used in satellite networks based on the target application \cite{kodheli2020satellite}. As depicted in Fig. \ref{top}, in a star topology, satellites are connected to a central node that controls their interconnections. In contrast, in a mesh setup, all satellites are directly connected \cite{yoon2019design}. Moreover, in line topology, the satellites are communicating with their neighbors only, following a line structure, as shown in Fig \ref{top}. Among these topologies, the star is by far the most popular for master-slave networks since it reduces the chances of network failures. However, mesh topology has more degree of freedom and less latency at the cost of more complexity because it enables more SSLL. Apart from the topologies, it is crucial to analyze the link design for both RF and optical-based SSLL to ensure sufficient connectivity and cooperation between the satellites.

\subsection{Satellite-to-Satellite Links at Same Layer (SSLL)}

Scientists from NASA, ESA, and DARPA  studied both intra- and inter-layer P2P satellite links, for over a decade. A.C. Clarke introduced the concept of satellite-to-satellite links  in 1945 \cite{evans2011}. Afterwards, SSLL became commonly used in satellite networks to offer cost-effective communication services. In contrast to the satellite-to-ground link, which is a duplex link, SSLL are mainly simplex links where the path length is measured by the LoS distance between any two satellites \cite{7964683}. In current systems, SSLL can be established by using either RF or FSO technologies \cite{nasa_2020}. In the following, we discuss the link budget analysis for both RF and optical SSLL.

\subsubsection{RF Link Budget}
In satellite communications, RF SSLL are the most widely used communication links because of their reliability and flexible implementation. Before calculating the link budget, it is essential to know the functional modulation and coding schemes used in RF-based links. Mainly, coherent systems such as Binary Phase Shift Keying (BPSK) are more desirable due to their lower power requirements to achieve a given throughput and bit error rate (BER). Nevertheless, the coherence capability produces delays as it takes time to lock the transmitted signal in the receiver terminal. Unlike coherent systems, non-coherent systems such as Frequency Shift Keying (FSK) require more transmitting power to achieve the same throughput and BER with less delay. Another popular modulation scheme for RF-based SSLL is Quadrature Phase Shift Keying (QPSK), which provides twice the bandwidth than a typical BPSK. QPSK, however, suffers from phase distortion because of the channel values, leading to system degradation, which is often solved using differential PSK in order to improve the overall spectral efficiency through striking a trade-off between power requirements and spectral efficiency \cite{radhakrishnan2016survey}.

For a given modulation scheme and under a non-coding assumption, the parameters used in calculating the link budget for RF-based SSLL can be described as a function of the satellite transmit power ($P_t$), the distance between satellites ($d$), achievable data rate ($R_b$), operating wavelength ($\lambda$), and diameter of the transmit antenna's aperture ($D$). For simplicity, the radiation of the transmitting antenna is assumed to be isotropic, where the radiation intensity is the same in all directions. Therefore, the gain of the transmitter and receiver antennas $G_t$  and $G_r$ can be calculated as follows:
\begin{equation}
\label{g}
G_t = G_r = \frac{4 \pi A}{\lambda^2},
\end{equation}
where $A=\frac{\pi D^2}{4}$ is the aperture of the antenna.
Besides the gain of the transmitter and receiver antennas, path loss $L_p$ is critical in the analysis and design of SSLL. Such pathloss can be calculated at the receiver antenna as follows
\begin{equation}
L_p =  {\left(\frac{4 \pi d}{\lambda}\right)}^{2},
\end{equation}\label{eq: fspl}
Based on the path loss, the received power is calculated as,
\begin{equation}
P_r = \frac {P_t G_t G_r}{L_p}.
\end{equation}\label{fspm}

To determine whether the received power is sufficient to establish a satellite-to-satellite link or not, we need to find the required signal-to-noise-ratio (SNR),  assuming that the noise is additive white Gaussian noise (AWGN). Such noise mimics the random processes effect in space, where the only communication impairment is the white noise. Besides, the required SNR primarily depends on the used modulation scheme and the target bit error probability ($P_b$) \cite{gibalina2017optical}. For instance, if the modulation scheme is BPSK, then the SNR required to achieve $P_b$  for the RF-based SSLL can be written as
\begin{equation}
\label{preq}
\gamma_{req} = \frac{E_b}{N_o}=  \frac{P_{r}}{k T R_b B},
\end{equation}
where $\frac{E_b}{N_o}$ is the bit-energy per noise spectral-density, $B$ is the bandwidth in Hertz, $k= 1.38\times10^{-23}$ is the Boltzmann constant, and $T=300$K is the absolute temperature \cite{sklar2001digital}. Hence, $P_b$ is calculated as
\begin{equation}
P_b = \frac{1}{2} \text{erfc}(\sqrt{\gamma_{req}}),
\end{equation}\label{pb}
where $\text{erfc}(\cdot)$ is the complimentary error function.

\begin{table}[ht]
\centering
\caption{Parameters for RF-based link budget calculation} \label{RFpar}
\begin{tabular}{|c|c|}
    \hline
    \textbf{Parameter} & \textbf{Value} \\ \hline\hline
    Transmitted power $P_t$ &2 W \\
    Satellite antenna gain $G_t$ \& $G_r$ & 0 dBi  \\
    Data rate $R_b$ & 1 Mbps \\
    Bandwidth $B$ & 0.5 MHz \\
    Antenna aperture area $A$ & 7.84 $\text{cm}^2$  \\
    Absolute temperature $T$ &300 K \\
    \hline \hline
\end{tabular}
\end{table}

\begin{figure}[h]
    \centering
    \includegraphics[width=\linewidth]{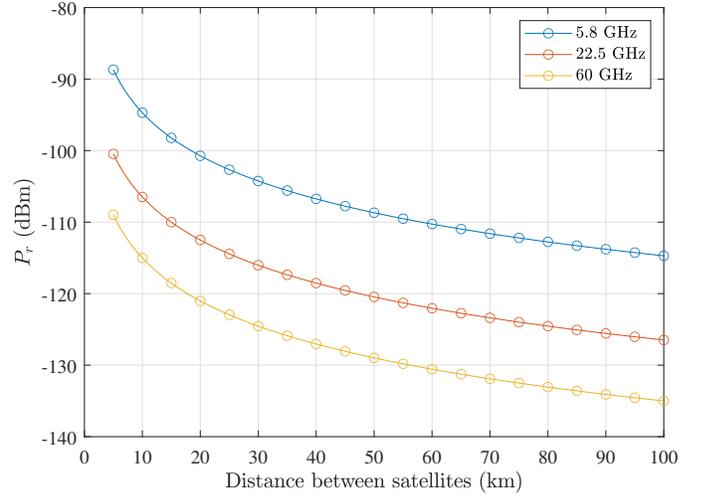}
    \caption{The received power for RF-based SSLL with varying link distances.}
     \label{rfpr}
\end{figure}

\begin{figure}[h]
    \centering
    \includegraphics[width=\linewidth]{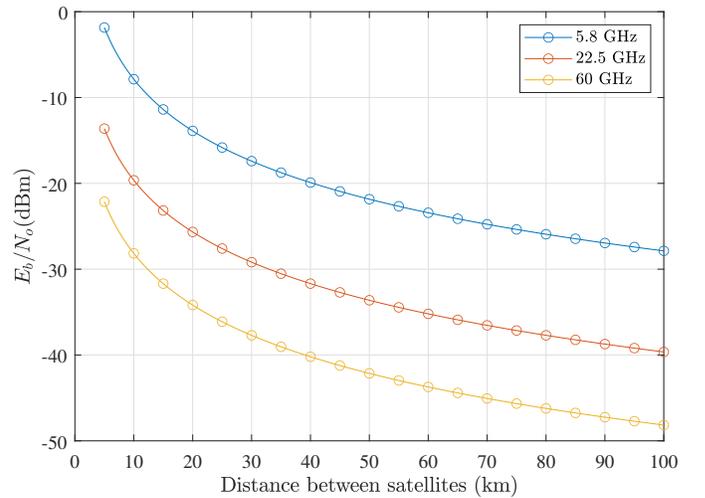}
    \caption{The energy-per-bit to noise spectral density for RF-based SSLL with varying link distances.}
     \label{rfsnr}
\end{figure}

We next give some numerical insights that highlight the above link-budget characterization. Consider RF-based SSLL among satellites orbiting in LEO. We first analyze the impact of distance and operating frequency on the received power and SNR. By varying the distance between satellites and the operating frequency of their interconnections, the received power is then calculated based on the above equations. Table \ref{RFpar} summarizes the parameters for calculating the RF-based link budget. From Fig. \ref{rfpr}, we observe that the received power is inversely proportional to the distance between satellites and the frequency. At the same distance, SSLL operating at a lower frequency results in a higher received power. This is mainly due to the frequency-dependent path loss, i.e., since the path loss increases at higher frequencies, the level of the received power decreases. On the basis of the International Telecommunication Union (ITU) recommendations, if we consider 22.5 GHz of frequency to establish SSLL, then -125 dBm power is received for a 100 km link. Note that SSLL with lower frequencies and distances have better energy-per-bit to noise spectral density with fixing the gain of the transmitted and received antennas. In Fig.~\ref{rfsnr}, we show the energy-per-bit to noise spectral density as a function of link distance. For instance, the energy-per-bit to noise spectral density values range between -2 and 19 dBm at 5 km for 60 GHz and 5.8 GHz, respectively. However, these values drop down to -48 and -28 dBm at 100 km.

\subsubsection{Optical Link Budget}
Another promising solution for establishing SSLL is using FSO, as it can offer superior data-rate compared to RF. Moreover, unlike RF communication, FSO systems are easily expandable, light in weight,
compact, and easily deployable. Even in terms of bandwidth, the permissible bandwidth can reach up to 20$\%$ of the carrier frequency in RF systems; however, the utilized bandwidth at an optical frequency is much higher even when it is taken to be 1$\%$ of the carrier frequency \cite{kaushal2016optical}. Nevertheless, high-speed optical links require a high directive beam that suffers from ATP challenges, as mentioned earlier, and hence, restricted to enable short-range SSLL. One possible solution to counter the ATP issue is  using photon-counting detector arrays at the receiver that improves the signal acquisition for long-range FSO communication \cite{Bashir2020, Bashir2020a, Bashir2020b}.

FSO communication supports various binary and high-level modulation schemes with different levels of power and bandwidth efficiency for SSLL \cite{kaushal2016optical}. The most widely adopted modulation format for optical SSLL is non-return-to-zero On-Off Keying (OOK-NRZ) due to its easy implementation, robustness, bandwidth efficiency, and direction detection facilitation. However, it imposes the constraint of an adaptive threshold for getting the best results \cite{6923082}.
On the other hand, M-Pulse Position Modulation (M-PPM) scheme does not require an adaptive threshold, offering better average-power efficiency, which in turn makes it a suotable choice for deep-space communications \cite{hemmati2006deep}. However, in case of limited bandwidth systems, increasing $M$ would cause the bandwidth efficiency to be substandard, and hence, high-level schemes are more favorable. Besides M-PPM, optical sub-carrier intensity modulation (SIM) does not require an adaptive threshold as well. Furthermore, it provides more bandwidth efficiency, less complicated design, and better bit error rate (BER) than the M-PPM scheme. On the contrary, the SIM scheme's major disadvantage is the inferior power efficiency as compared to OOK-NRZ and M-PPM \cite{chatzidiamantis2011adaptive}.
According to \cite{10.1117/12.698646}, homodyne BPSK is a recommended coherent modulation scheme for SSLL because of its better communication and tracking sensitivity. Moreover, it also gives complete protection from solar background noise. Another good candidate is the differential phase-shift keying (DPSK) modulation scheme. It considerably reduces power requirements and enhances spectral efficiency than OOK-NRZ. However, it is complex to design and hence expensive to implement \cite{atia1999demonstration}.
\begin{figure}[h]
    \centering
    \includegraphics[width=\linewidth]{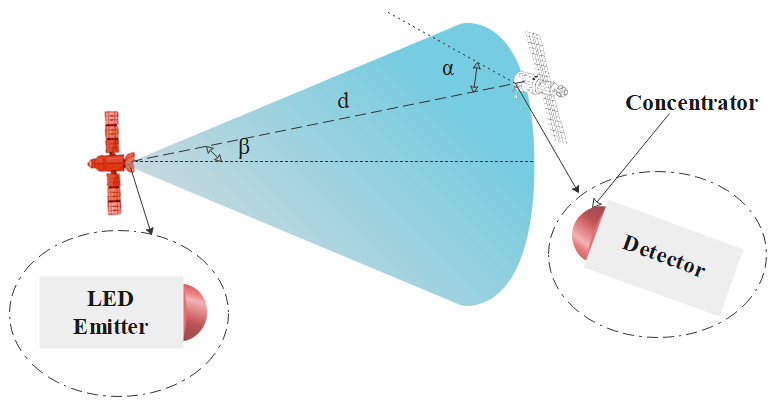}
    \caption{FSO-based LoS satellite-to-satellite link.}
     \label{opt}
\end{figure}

To calculate the optical link budget,  we next consider light-emitting diodes (LEDs) as transmitters and photodetectors as receivers. The LEDs are assumed to use the OOK-NRZ modulation scheme for enabling an optical SSLL. At the receiver, the detector's choice depends on various factors, including cost, power level, the wavelength range of the incident light, and the detector amplifier bandwidth.  We refer the interested readers to \cite{wood2010using, kahn1997wireless,kharraz2013performance} for a detailed overview of the types of photodetectors.

The generic LoS optical SSLL is illustrated in Fig. \ref{opt} where $d$ is the distance between satellites, $\alpha$ is the angle of incidence with respect to the receiver axis, and $\beta$ is the viewing angle (irradiance angle) that describes the focus of the LED emitting beam. In LoS optical links, the channel DC gain $H(0)$ is calculated as
\begin{equation}
\label{h}
H(0)=
\begin{cases}
\frac{(m+1)}{2 \pi d^2} A_o \cos^m{(\beta)} T_f g(\alpha) \cos{(\alpha)}, & :0 \leq \alpha \leq \alpha_c \\
0, &: \alpha> \alpha_c
\end{cases},
\end{equation}
where $m$ represents the order of Lambertian emission (i.e., a quantity that expresses the radiation characteristics shape), $T_f$ is the filter transmission coefficient, $g(\alpha)$ is the concentrator gain, and $A_o$ is the detector active area. The value of $m$  is related to the receiver field of view (FoV) concentrator semi-angle $\alpha_c$ at half illuminance of an LED $\Phi_{1/2}$ as $m= \frac{- \ln{2}}{\ln(\cos{\Phi_{1/2}})}$. Following the analysis in \cite{amanor2018intersatellite} and \cite{lee2009performance}, an extra concentrator gain is achieved by utilizing a hemispherical lens with internal refractive index $n$ as
\begin{equation}
\label{galph}
g(\alpha)=
\begin{cases}
\frac{n^2}{\sin{\alpha_c}}  & :0 \leq \alpha \leq \alpha_c \\
0, &: \alpha> \alpha_c
\end{cases}.
\end{equation}
Hence, the received optical power ($P_{r_o}$) can be expressed as
\begin{equation}
\label{pro}
P_{r_o} = H(0) P_t,
\end{equation}
At the receiver side, the electrical signal component can be expressed by
\begin{equation}
\label{s}
S = (\xi P_{r_o})^2
\end{equation}
\noindent where $\xi$ is the photodetector responsivity. Therefore, the required SNR at the receiver side can be determined given that the total noise variance $N$ is the sum of noise variances (shot noise $\sigma_{s}^2$ and thermal noise $\sigma_{t}^2$), as
\begin{equation}\label{opticalsnr}
 \gamma_{req} = \frac{E_b}{N_o}= \frac{\left[\xi H(0) P_t\right]^2}{N} \frac{B}{R_b}.
\end{equation}
\noindent Further evaluation of $\sigma_{s}^2$ and $\sigma_{t}^2$ can be found in \cite{amanor2018intersatellite}. Based on (\ref{opticalsnr}), $P_b$ for OOK scheme can be calculated as

\begin{equation}
P_b = \frac{1}{2} \text{erfc}\left(\frac{1}{2\sqrt{2}}\sqrt{\gamma_{req}}\right).
\end{equation}\label{pbopt}

\begin{table}[ht]
	\centering
	\caption{Parameters for the optical link budget calculation.} \label{optpar}
	\begin{tabular}{|c|c|}
		\hline
		\textbf{Parameter} & \textbf{Value} \\ \hline\hline
		Transmitted power $P_t$ &2 W \\
		Semi-angle at half power $\Phi_{1/2}$ & $30^\circ$   \\
		Incidence angle $\alpha$ & $30^\circ$ \\
		Irradiance angle $\beta$ & $15^\circ$ \\
		Detector responsivity $\xi$ &0.51 \\
		Refractive index of lens $n$ &1.5 \\
		Data rate $R_b$ & 1 Mbps \\
		Bandwidth $B$ & 0.5 MHz \\
		Detector active area $A_o$ & 7.84 $\text{cm}^2$  \\
		Absolute temperature $T$ &300 K \\
		Filter transmission coefficient $T_f$ & 1.0 \\
		LED wavelength $\lambda$ & 656.2808 nm \\
		\hline \hline
	\end{tabular}
\end{table}

\begin{figure}[h]
	\centering
	\includegraphics[width=\linewidth]{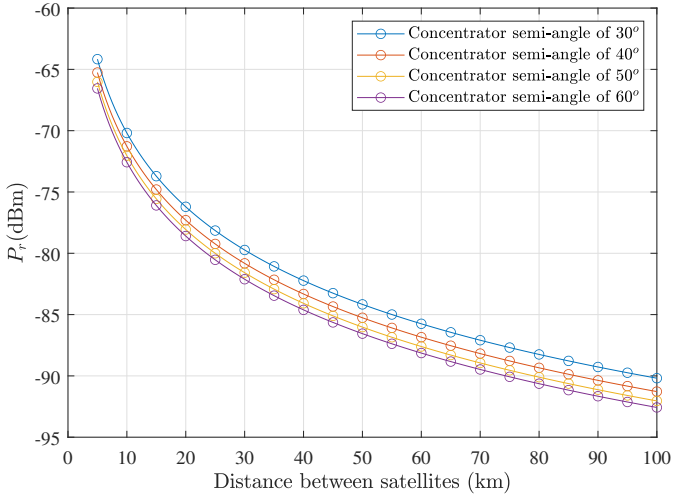}
	\caption{The received power for different optical SSLL distances.}
	\label{optpr}
\end{figure}

\begin{figure}[h]
	\centering
	\includegraphics[width=\linewidth]{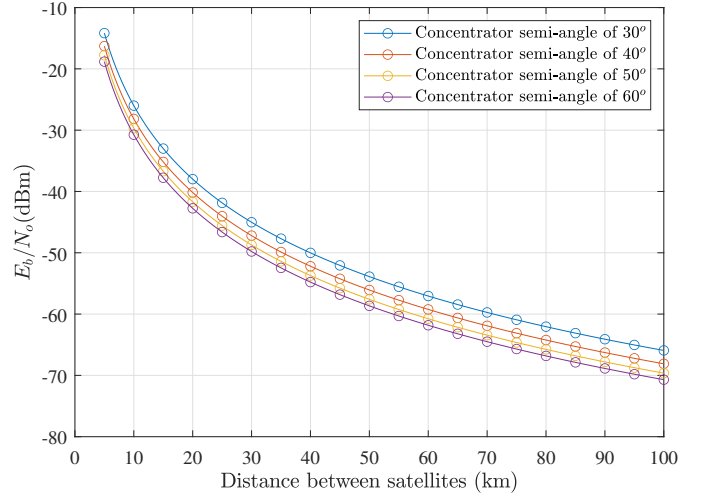}
	\caption{The energy-per-bit to noise spectral density for different optical SSLL distances.}
	\label{optsnr}
\end{figure}

We now present a numerical link budget illustration by considering a setup similar to the RF setup described earlier, where the satellites orbit in LEO but with optical SSLL. The parameters used for the simulations are mainly taken from \cite{amanor2018intersatellite} and are listed in Table \ref{optpar}. In Fig.~\ref{optpr}, we plot the received power as a function of the concentrator FoV semi-angle. As expected, Fig. \ref{optpr} illustrates that as the distance between the satellites increases, the received power decreases. Also,  in the case of a smaller concentrator angle, slightly more power is received. Furthermore, in comparison with the RF case, the received power using the optical technology is higher. For example, at 5 km, the optical received power is approximately -50 dBm; however, it swings between -70 and -90 dBm in the RF scenario. Moreover, Fig. \ref{optsnr} presents the influence of the concentrator FoV semi-angles on the energy-per-bit to noise spectral density for different distances, where the performance degrades with increasing the FoV of detectors and the distance between satellites.

\subsection{Satellite-to-Satellite Links at Different Layers (SSLD)}
 Despite the fact that a single layer satellite network designed by GEO, MEO, or LEO with P2P SSLL can offer multi-media services to some degrees, many restrictions can affect the performance of such a single layer satellite network. For instance, a high accumulated delay is present in large constellations due to multi-hops, and low stability is expected because of the single-layer satellite network with planar topologies. Moreover, repeated handovers lead to an increase in the probability of network routing and re-routing, creating congestions \cite{wang2006analysis}. All the restrictions above harden the establishment and maintenance of a single-layer satellite network.

Therefore, many studies on satellite-to-satellite links at different layers (SSLD) exist in the literature. For instance, in 1997, \cite{kimura1997double} proposed the earliest two-layer satellite constellation comprising of MEO and LEO satellites. The architecture in \cite{kimura1997double}  consists of both SSLL (among MEO satellites) and SSLD (between LEO and MEO satellites). Consequently,  \cite{lee2000satellite} proposed a similar two-layer MEO and LEO satellite network, which included SSLL in each layer besides the SSLD. Their network was designed to transmit short distance-dependent services through SSLL, and relay long-distance traffics via MEO satellites using SSLD. \cite{akyildiz2002mlsr} introduces instead a more complex multilayer satellite network architecture consisting of GEO, MEO, and LEO satellites to improve capacity, reliability, and coverage of satellite communication networks.

To implement such a multilayer satellite network, Japan Aerospace Exploration Agency (JAXA) made various attempts to develop a space data relay network for the next generation of wireless communication systems.  Moreover, various other projects also tried to implement such multilayered satellite networks with SSDL. Most of the recent works prefer to use FSO for enabling satellite-to-satellite links at different layers. One such project is Optical Inter-orbit Communications Engineering Test Satellite (OICETS) ``Kirari'' by JAXA that uses optical P2P links between satellites at different orbits. Another similar project is ``ARTEMIS'' by ESA that also uses optical links between the satellites at different altitudes \cite{jono2006oicets,fujiwara2007optical,yamakawa2010jaxa,jono2007overview}.  Some other similar projects are Alphasat TDP1 Sentinel-1A that uses FSO to relay data from GEO to LEO \cite{TDP1} \cite{10.1117/12.2212744}.
Moreover, recently, reference \cite{GROVER2020164295} propose a 20 Gbit/s-40 GHz OFDM based LEO-GEO optical system using 4-QAM modulation. Similarly, \cite{li2010novel} presents a novel two-layer satellite LEO/MEO network with optical links. On the basis of the link quality, \cite{zhou2007novel} introduces a novel QoS routing protocol for LEO and MEO satellite networks. Furthermore, Yan et al. discuss the topology analysis of two-layer links in LEO/MEO  satellite networks \cite{10.1007/978-981-10-7877-4_13}. FSO communication provides a promising solution to enable satellite-to-satellite links at different altitudes because the radiated light beam is not affected by the turbulence. However, FSO requires efficient ATP mechanisms to provide reliable and stable links.



\begin{table*}
	\caption{List of various projects on HAPs.}\label{HAP_projects}
	\centering	
	\small
	\begin{tabular}{|p{2.0cm}|p{2cm}|p{1.5cm}|p{1.8cm}|p{2.2cm}|p{6.0cm}| }
		\hline
		\hline
		\textbf{Project}  & \textbf{Type} & \textbf{Technology} & \textbf{Link Type}  & \textbf{Organization} & \textbf{Description} \\
		\hline
		\hline
		SHARP \cite{jullsharp} & Aerodynamic &Microwave &HAP-Ground &Communications Research Centre (CRC) & It goes to prove successful one-hour communication flight time.
		\\
		\hline
		Pathfinder, Centurion, and  Helios \cite{pan_2020} & Aerodynamic &RF &HAP-Ground  &  NASA  & This project consists of a solar powered aerodynamic HAP providing high-definition TV (HDTV) transmissions and 3G communication services.\\
		\hline
		SkyNet \cite{karapantazis2005broadband,d2016high,aragon2008high} &Aerostatic- (Airship) &RF &HAP-Ground &JAXA & SkyNet promotes future high-speed wireless communications by using a 200 m length airship that can operate for up to 3 years.\\
		\hline
		CAPANINA \cite{capanina} & Aerostatic- (Balloon) &Optical and RF & HAP-Ground & University of York & This project provides enhance broadband access for both urban and rural communities in Europe, demonstrating data transmission of 1.25 Gbps.\\
		\hline
		X-station \cite{stratxx_2020}& Aerostatic- (Airship) &RF &HAP-Ground &StratXX & X-station airship can stay in the air for around an year providing various communication services, such as TV and radio broadcasting, mobile telephony, VoIP, remote sensing, and local GPS.
		\\
		\hline
		
		Elevate \cite{zero} & Aerostatic- (Balloon) &RF &HAP-Ground & Zero 2 Infinity & Elevate balloons can lift payloads up to 100 kg to test and validate novel technologies in the stratosphere.
		\\
		\hline
		Loon \cite{loon} & Aerostatic- (Balloon) &Optical & HAP-Ground and IHAP  & Alphabet Inc.  &  The aim of this project is to connect people globally using a network of HAPs with each balloon having 40 km of coverage radius. The balloons in this project can stay in the air for 223 days.\\
		\hline
		Zephyr S \cite{zephyr} & Aerodynamic &RF & HAP-Ground & Airbus & Project Zephyr S can lift a payload of up to 12 kg and can flight continuously for around 100 days, aiming to connect the people in underserved areas, achieving 100 Mbps.\\
		\hline
		Aquila \cite{cox_2020} & Aerodynamic &RF &HAP-Ground & Facebook & Similar to Zephyr S, the goal of Aquila was to provide broadband coverage in remote areas.\\
		\hline
		
		Stratobus \cite{stratobus} & Aerostatic- (Airship) &Optical & HAP-Ground and IHAP & Thales Alenia Space & Unlike other HAPs, Stratobus can support heavy payload, i.e.,  up to 450 kg and stay almost static in the stratosphere for a longer time (up to 5 years), providing 4G/5G communication services.\\
		\hline
		
		HAWK30 \cite{mobile_2020} & Aerodynamic &mmWave & HAP-Ground & SoftBank Corp. & This project consists of HAPs with each having 100 km of coverage, aiming to ground users, UAVs, IoT devices. \\
		\hline
		
		PHASA-35 \cite{prismatic_2020} &Aerodynamic &RF &HAP-Ground & Prismatic & Project PHASA-35 can support up to 35 kg of payload and can fly continuously for an year to  provide  5G communication services.\\
		\hline
	\end{tabular}
\end{table*}

\section{P2P Links in HAP Networks}
Unlike the satellites, HAPs operate at a much lower altitude, i.e., around 20 km in the stratosphere above the earth's surface. The HAPs can provide ubiquitous connectivity in the operation area since they can stay quasi-static in the air \cite{tozer2001high, mohammed2011role, grace2011broadband, grace2005integrating}. Numerous research projects use HAPs to enable connectivity, especially in rural areas or in disaster-affected regions. One such example is the Google Loon project, which aims to provide Internet access in underserved areas. Table \ref{HAP_projects} presents numerous HAPs projects that aim to develop aerial base stations. Recently, HAPs-based wireless connectivity solutions are promising due to the advances in the development of lightweight materials and efficient solar panels that increase the lifetime of HAPs and reduces the cost. Accordingly, a set of inter-connected HAPs can be a transpiring solution to provide Internet access and remote sensing in a broad region. Therefore, it is interesting to discuss potential connectivity solutions among HAPs that can lead to extended coverage and perform backhauling.

\subsection{HAP-to-HAP Links (HHL)}
Early studies on establishing HAP-to-HAP Links (HHL) and HAP backhauling mainly focus on radio communications. However, implementing RF links either for inter-HAP communication or backhauling is not suitable for multiple reasons, e.g., such links require high bandwidth and high transmit power for long-range communication \cite{arum2020review}. Besides, wireless communication links at a higher RF frequency band are severely affected by environmental impediments, such as rain attenuation. Irrespective of these challenges, various works studied RF-based HHL and backhaul links \cite{thornton2008wimax, yang2010channel,  nauman2017system, abdulrazak2017stratospheric, Shibata2020, Zhao2020, Popoola2020}. For instance, \cite{thornton2008wimax} proposes a backhaul link between the HAP with WiMAX payload and the customer premises on the ground. Consequently, \cite{nauman2017system} investigates digital video broadcasting protocol (DVB-S2) for the backhauling to the ground station by using HAPs, which shows that the BER is low compared to WiMAX at lower SNR. \cite{abdulrazak2017stratospheric} highlights the effects of weather conditions on the performance of HAPs backhaul links. Moreover, recently, \cite{Shibata2020} optimizes the cell configuration for a high-speed HAPs system by using a genetic algorithm that also tries to minimize the total power consumption.

Besides HAPs backhauling, interconnecting the HAPs require high-speed communication links. Therefore, unlike the HAP-to-ground links, which mainly uses RF communication, establishing inter-HAP links prefer to use FSO communication  \cite{lun2017tv, alzenad2018fso}. The FSO links are vulnerable to weather conditions, such as clouds and fog. However, the HAPs are operating above the clouds; thus, FSO links are less affected at such an altitude. For example, \cite{Katimertzoglou2007} proposes a 500 km inter-HAP FSO link at 20 km of altitude, achieving 384 Mbps of data rate with $10^{-6}$ BER. Likewise, \cite{akbar2015ber} performs BER analysis for FSO-based inter-HAP links in the presence of atmospheric turbulence, where the BER increases with an increase in the scintillation index and link distance. In order to evaluate the performance of FSO-based HHL, it is important to develop accurate channel models that account for various losses such as geometrical loss, turbulence, inhomogeneous temperature, and pointing error. Geometrical loss mainly occurs due to the spreading of light resulting in less power collected at the receiver. On basis of the path length $d$, radius of the receiver aperture $r$, and divergence angle $\alpha$, the geometrical loss can be represented as
\begin{equation}\label{eq:fsogl}
L_g = \frac{4 \pi r^2}{\pi (\alpha d)^2}.
\end{equation}
Similarly, the estimation of turbulence loss requires to measure the turbulence strength with changing refractive index parameter $n^2(h)$ at various altitudes. Various empirical models, such as Hufnagel-Valley (H-V) model are used to estimate $n^2(h)$. On the basis of (H-V) model, $n^2(h)$ as a function of altitude ($h$) is measured as
\begin{eqnarray}\label{eq:tur}
n^2(h) &=& 0.00594\left(\frac{\nu}{27}\right)^2(10^{-5}h)^{10}\exp\left(\frac{-h}{1000}\right)\\
& & +2.7\times 10^{-16} \exp{\left(\frac{-h}{1500}\right)}+K\exp\left(\frac{-h}{100}\right), \nonumber
\end{eqnarray}
where $\nu$ is the wind speed and $K= 1.7 \times 10^{-14} \text{m}^{-2/3}$ is constant. Based on \eqref{eq:tur}, the turbulance loss in dB's is calculated as
\begin{equation}
L_t = 2\sqrt{23.17 \left(\frac{2\pi}{\lambda} 10^9\right)^{7/6} n^2(h) d^{11/6}}.
\end{equation}
Additionally, the pointing loss occurs due to numeorus reasons such as wind, jitter, turbulence, and vibration of HAPs. The pointing error can result in a link failure or reduces the amount of power received at the receiver resulting in a high BER.
Therefore, it is crucial to model the pointing error both in azimuth and elevation. There are various statistical distributions in the literature to model the pointing error for FSO communication, such as Rayleigh distribution \cite{Farid2007}, Hoyt distribution \cite{gappmair2011ook}, Rician distribution \cite{yang2014free}, and Beckmann distribution \cite{AlQuwaiee2016}.
In case when the pointing error is modeled as Gaussian
distribution, the radial error angle $e = \sqrt{\theta^2 + \phi^2}$ is the function of elevation ($\theta$) and azimuth ($\phi$) angles. Considering that  $\theta$ and  $\phi$ are zero-mean i.i.d processes with variance $\sigma$, then the pointing error follows Rician distribution as follows
\begin{equation}\label{eq:pointing}
f(\theta, \beta) =\frac{\theta}{\sigma^2}\exp\left(-\frac{\theta^2+\beta^2}{2 \sigma^2}\right)I_0\left(\frac{\theta \beta}{\sigma^2}\right),
\end{equation}
where $\beta$ is the angle bias error from the center and $I_0(\cdot)$ is the zeroth-order Bessel function. In case when $\beta = 0$, \eqref{eq:pointing} leads to Rayleigh distribution function, given as
\begin{equation}
f(\theta) =\frac{\theta}{\sigma^2}\exp\left(-\frac{\theta^2}{2 \sigma^2}\right).
\end{equation}
The pointing error for FSO-based inter-HAP links can be mitigated by increasing the receiver FoV, using multiple beam transmissions, hybrid RF/FSO, and adaptive optics \cite{Kaushal2017}. In the literature, various statistical channel models can be found that models the propagation characteristics of FSO communication.
For example, \cite{Habash2001} propose a gamma-gamma distribution for a laser link in the presence of turbulence. \cite{majumdar2005} uses log-normal distribution to model the FSO links with fluctuations. 
These statistical fading models can estimate the scintillation index for FSO links and help in analyzing these links. For example, the log-normal distribution estimates well the weak turbulence; however, it underestimates the distribution's tails and peaks. In contrast, exponential channel distribution fits well for a strong turbulence region but is not consistent for weak turbulence. Nevertheless, the gamma-gamma channel model works well for both weak and strong turbulence regimes \cite{Habash2001}. Similarly, Malaga distribution also fits well for a wider range of turbulence effects where log-normal and gamma-gamma distributions are its special cases. In the case of a gamma-gamma channel model,  the probability distribution function (PDF) for the irradiance $I_r$ can be written as
\begin{equation}
f_{I_r}(I) = \frac{2 (\bar{\alpha} \bar{\beta} )^{\frac{\bar{\alpha} + \bar{\beta}}{2}}}{\Gamma (\bar{\alpha}) \Gamma (\bar{\beta})} I^{\frac{\\bar{alpha} + \bar{\beta}}{2}} J_{\bar{\alpha} - \bar{\beta}} \left(2 \sqrt{\bar{\alpha} \bar{\beta}} I\right)
\end{equation}
where $\bar{\alpha}$ and $\bar{\beta}$ are the fading parameters for turbulence, $\Gamma(\cdot)$ is the gamma function, and $J (\cdot)$ is the second order modified Bessel function. Based on the values of $\bar{\alpha}$ and $\bar{\beta}$, the scintillation index for gamma-gamma model can be written as
\begin{equation}
\sigma_I = \frac{1}{\bar{\alpha}} + \frac{1}{\bar{\beta}} + \frac{1}{\bar{\alpha} \bar{\beta}}  
\end{equation}
Note that the effect of turbulence can be mitigated by using aperture averaging, i.e., increasing the aperture size reduces the fluctuations leading to a lower scintillation index \cite{AlQuwaiee2016}. The interested readers are referred to \cite{Trichili2020} for various FSO channel models that can be used for establishing inter-HAP links.

In the presence of the impediments mentioned above, researchers have studied the performance HAPs regarding coverage and capacity. Nevertheless, most of the existing works study HAP-to-ground links using geometrical and statistical models \cite{Guan2020}. For instance, \cite{palma2010wimax} investigates BER performance for hybrid WiMAX and HAP-based connectivity solutions for ground users. \cite{zakia2017capacity} performs the capacity analysis for a MIMO-based communication link between the HAP and a high-speed train, which shows that although there is a strong LoS component, the channel is still ill-conditioned. Similarly, \cite{horwath2007experimental} designs HAPs-based backhaul link using FSO in the presence of turbulence, achieving 1.25 Gbps with BER of less than $10^{-9}$. Consequently, \cite{michailidis2010three} studies a 3D channel model to see the impact of distance among antennas in a MIMO-HAP system, where the channel is affected by the distribution of scatters, array configuration, and Doppler spread. Moreover, \cite{sudheesh2018effect} investigates interference for ground users with two HAPs, showing that better performance is achieved if the users are spatially well separated. In \cite{thornton2003optimizing}, the authors improve the capacity of HAP systems by using mmWave frequencies. \cite{thornton2003optimizing} also evaluates ground users' capacity regarding the angular separation between the ground users and HAPs. Furthermore, \cite{thornton2003optimizing} analyze the coverage of HAPs operating at 48 GHz and 28 GHz frequencies discussing various crucial system parameters, including beam type and frequency reuse for cell planning. \cite{Xu2017} focuses on the deployment of HAPs to characterize the HAP-to-ground link in terms of path loss and maximizes the on-ground coverage.

\begin{figure*}[h]
    \centering
    \includegraphics[width=0.9\linewidth]{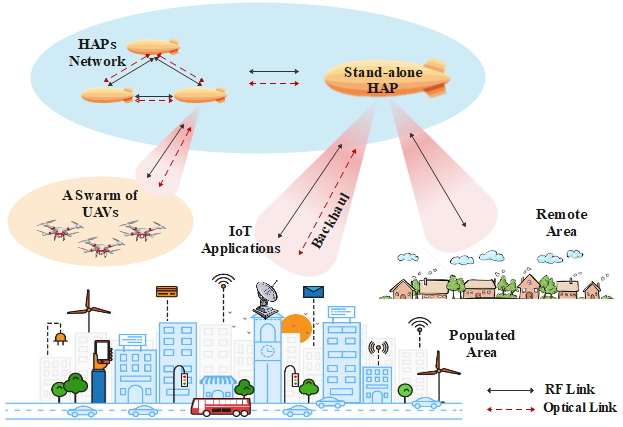}
    \caption{An architecture of HAPs network with P2P HAP-to-HAP and backhauling links.}
     \label{HAP}
\end{figure*}

Moreover, \cite{datsikas2010serial} investigates the use of relays in the presence of turbulence and pointing errors for multi-hop FSO that can be used for establishing inter-HAP links. \cite{datsikas2010serial} analyzes amplify-and-forward relaying with channel state information and fixed-gain relays regarding signal-to-interference-plus-noise ratio (SINR) and coverage probability. Consequently, \cite{sharma2016high} derives the closed-form expression of BER and channel capacity showing the effects of pointing errors and beam wandering for FSO-based inter-HAP links. Michailidis et al. further investigates hybrid triple-hop RF-FSO-RF links for HAPs based communication systems where the two HAPs are connected through FSO while the HAP-to-ground link is RF \cite{michailidis2018outage}. Fig. \ref{HAP} illustrates such a hybrid RF-FSO architecture where FSO can be used in good weather conditions to achieve higher data rates while RF can be utilized in bad weather conditions and in the absence of LoS.

\subsection{Handover between HAPs}
The HAPs in the stratospheric atmosphere can be affected by the airflow, resulting in a different footprint on the ground. Therefore, it is crucial to design handover schemes for the ground users to maintain the communication link. Handover in HAP networks is the process of transferring the communication link between cells to avoid the channel's instability. This process usually occurs when there are massive differences between cell sizes in HAP extended coverage scenarios \cite{arum2020review}. Many works in the literature discuss handover schemes for a stand-alone HAP or between HAP networks \cite{lim2002adaptive,katzis2010inter, li2010directional, li2010novel, li2011cooperative, handover2016}. In \cite{li2010directional,li2010novel,li2011cooperative}, the authors focus on minimizing the traffic difference between cells during the data transfer, considering the HAP travel direction, the adaptive modulation, and cells cooperation, respectively.
On the other hand, Lim et al. suggest an adaptive soft handover algorithm using both the platform's downlink output power and individual base stations in \cite{lim2002adaptive}. In \cite{katzis2010inter}, the authors discuss the influence of platform movement on handover. Moreover, a handover decision algorithm based on prediction, using the time series analysis model with an adaptive threshold, is designed in \cite{handover2016}. We wish to finally mention that most the link budget illustrations of P2P links in satellite networks discussed in the previous section also apply to HAP networks, and so we choose not to explicitly describe them in the text for conciseness.


\section{Integrated Satellite-HAP Communication Networks}
6G wireless communication systems envision to provide broadband services in underserved areas with reasonable costs. Satellite networks are one possible enabler of such a vision due to their large footprints and their capabilities to provide ubiquitous coverage to remote areas. Recently, mega-constellations of small satellites in LEO gain interest in academia and industry to enable broadband services worldwide \cite{seitzer2020mega}.  Moreover, the development of integrated satellite-HAPs-LAPs networks can further improve the coverage, reliability, and scalability of 6G wireless communication systems \cite{pace2010satellite,Pan2020, Ye2020}.
A potential integrated spatial network consists of spatial nodes at the same or different altitudes connected via either RF or optical links. For example, satellite networks can provide RF/optical backhauling for HAPs and LAPs.

Recently, various research works are devoted to the vision of integrated spatial networks. For example, \cite{Zhang2017} proposes an integrated spatial and terrestrial system consisting of satellites with mounted BSs, UAVs, and ground vehicles. Their solution is based on densification to increase the network capacity in the demand area. However, the proposed architecture in \cite{Zhang2017} is a function of several challenges, such as interoperability, resource allocation, and network management for a highly dynamic environment. To this end, \cite{Zhou2018} develops SAGECELL, a software-defined integrated spatial/terrestrial moving cell solution. The SDN-based approach results in flexible resource allocation with centralized network management.
Moreover, \cite{Di2019} proposes an integrated satellite-terrestrial access network where the LEO-based small cells coordinate with small terrestrial cells to improve wireless access reliability and flexibility. However, this approach requires ultra-dense deployment of LEO satellites and also ignores HAPs and LAPs. Zhu et al. propose a cloud-based integrated satellite-terrestrial network where both the satellite and ground BSs are connected to a common baseband processing system that performs interference mitigation, cooperative transmission, and resource management \cite{Zhu2019}.

Unlike the works mentioned above, \cite{Yao2018} introduces a heterogeneous spatial network consisting of satellites, HAPs, and LAPs. The backbone network entities are connected via laser links and the access network, allowing the user to enter the spatial network using microwave links. Several industrial projects have been launched to realize such an architecture. For example, Integrated Space Infrastructure for global Communication (ISICOM) \cite{Coralli2009 } and Transformational Satellite Communications System (TSAT) \cite{Pulliam2008} aim to provide global communication, covering oceans, ground, and space. Moreover, various works investigate the communication link between HAPs and satellites. For instance, \cite{perlot2008system} explores optical HAP-to-LEO links where the reliability of the link degrades at low elevation angles. Similarly,  \cite{vu2018performance} proposes a HAP-based relaying for FSO communication between the ground and LEO satellites. Thanks to the HAP-based relaying, it increases the power gain by 28 dB at BER of $10^{-9}$ \cite{vu2018performance}.


\section{Future research directions}
On the basis of the literature we reviewed, this section outlines numerous promising future research challenges for integrated spatial networks. Since the studies on these complex, large-scale spatial networks are still at initial stages, various problems need further investigation. In the following, we point out to some of these open research issues.

\subsection{Network Optimization}

Network optimization for an integrated spatial network is much more complicated than a stand-alone terrestrial or an aerial network because of the diverse characteristic of spatial nodes at each layer. Therefore, novel optimization techniques are required to consider various network characteristics, such as cost, mobility, energy efficiency, spectrum efficiency, and user experience. Recently, the use of artificial intelligence is gaining interest in optimizing such large-scale networks. For instance, \cite{Zappone2018} employs a deep neural network model to optimize wireless networks' energy consumption. Similarly, \cite{Asheralieva2017} uses reinforcement learning with a Bayesian network to maximize the throughput of a D2D network. Likewise, \cite{Mwanje2016} targets to improve mobility robustness using Q-learning for cellular networks. Recently, \cite{Kato2019} uses artificial intelligence to optimize integrated spatial and ground networks regarding traffic control, resource allocation, and security. However, the existing works on optimization for spatial networks remain relatively limited, and so advanced joint optimization techniques need to be developed to address various issues of spatial networks, such as cost, spectrum utilization, security, traffic offloading, and energy efficiency.

\subsection{Intelligent Offloading}
There has been a plethora of work on traffic offloading in different wireless networks, including satellite, UAVs, and terrestrial networks \cite{Di2019}. With the recent advancements in integrated spatial networks, new possibilities for traffic offloading arise. Nevertheless, resource management and coordinated traffic offloading in such an integrated network are more complicated than a standalone non-terrestrial or terrestrial network \cite{Zhou2019}. For example, satellite connections have large latency, which means low QoE compared to terrestrial links. Concurrently, satellite links are more appealing for continued services and seamless connectivity due to its wider footprint. Recently, \cite{Abderrahim2020} proposes a latency-aware scheme for traffic offloading in integrated satellite-terrestrial networks where the URLLC requirement is satisfied for traffic offloading to the terrestrial backhaul.
In contrast, eMBB data is offloaded to the satellites as eMBB traffic does not have always a stringent delay requirement. Moreover, intelligent traffic offloading in integrated spatial-terrestrial networks can be enabled using SDN technology that can separate the data and network plans \cite{Niephaus2019}. Also, based on link characteristics, such as cost, reliability, and capacity, multiple options can offload the data.  Therefore, it is interesting to investigate different traffic offloading schemes for integrated spatial-terrestrial networks to make optimum offloading decisions.

\subsection{Smart Platforms}
Intelligent reflecting surfaces, also known as smart surfaces (SS) have emerged as promising 6G wireless communication technology. These smart surfaces consist of flexible metamaterials that allow them to passively/actively reflect the received signals improving the communication channel's quality \cite{Rawan2020}. Considering numerous smart surfaces' opportunities, it is well-suited for the spatial platforms, including satellites, HAPs, and UAVs \cite{alfattani2020aerial}. For instance, \cite{tekbyk2020} proposes SS-assisted THz communication links for LEO satellite networks where SS improve the SNR of the received signal. Similarly, \cite{alfattani2020link} investigates the link budget analysis for communication in SS-assisted aerial platforms.
SS-assisted spatial platforms offer several advantages, including energy efficiency, improved coverage, and lower system complexity. Despite these benefits, the research on SS-assisted spatial platforms is in infancy and needs further investigation.

\subsection{Energy Efficiency}
The limited power supply of spatial platforms requires to use the on-board energy efficiently. Unlike terrestrial networks where most of the energy is consumed in communication, spatial networks are also affected by radiations, space/aerial environment, and different propagation channels \cite{Yang2016}. One way to reduce spatial platforms' power consumption is to design power amplifiers with a low peak-to-average power ratio (PAPR). Novel techniques such as non-orthogonal waveforms can be investigated to reduce the PAPR. Moreover, spatial platforms' energy consumption can also be reduced by using new networking technologies, such as SDN and NFV. In \cite{Zhang12017}, the authors reveal that significant energy gain can be accomplished for integrated spatial-terrestrial networks by splitting the control and data plans using SDN. Furthermore, energy harvesting techniques need to be explored to make spatial networks green and environment friendly.

\subsection{Novel Multiple Access Schemes}
Several multiple access schemes, such as space-division multiple access (SDMA) and non-orthogonal multiple access (NOMA), are promising for multiplexing in aerial networks. However, the gain of SDMA and NOMA is limited because they depend on environmental conditions. Therefore, \cite{Rahmati2019} introduces rate-splitting multiple access (RSMA), which has better spectral efficiency for an integrated network. In the context of integrated spatial-terrestrial networks, RSMA can be employed horizontally at one of the layers or vertically at each layer \cite{jaafar2020multiple}. The management of RSMA can be performed centrally (if a central controller manages a layer) or in a distributed fashion (if layers are separately managed). Nevertheless, the investigation of RSMA in such scenarios is missing in the literature and needs the researchers' attention.

\subsection{Distributed Spatial Networks}
The spatio-temporal variations of the flying platforms and their relative positioning are critical aspects of the ground-level communications metrics. While satellites move in pre-determined constellations which typically consist of complementary orbital planes \cite{saeed2019cubesat}, HAPs are relatively stationary within the stratospheric layer \cite{arum2020review}. LAPs (e.g., UAVs), on the other hand, are distributed platforms capable of dynamically adjusting their locations based on both the underlying ground-demand, and the heterogeneous nature of the wireless network; see \cite{hammouti2020optimal} and references therein. Automating LAPs positioning becomes, therefore, an important aspect of terrestrial-aerial networks design so as to improve the overall system quality-of-service. From an end-to-end system-level perspective, the provisioning of the spatio-temporal variations of the network (e.g., data traffic, user-locations, etc.) and the positioning of the aerial networks (e.g., UAVs dynamic positioning, satellite constellations design, HAPs placement, etc.) becomes crucial both to capture the instantaneous and the long-term network metrics, and to optimize the network parameters accordingly. A future research direction is, therefore, to enable the real-time operations of such distributed systems, mainly LAPs-to-LAPs and LAPs-to-ground, through the accurate modeling of the networks variations, and through invoking the proper online distributed optimization for real-time data processing.

\section{Conclusions}
Spatial networks are emerging as major enablers for next-generation wireless communications systems. Through their invigorating capabilities in providing connectivity solutions, improving remote areas' coverage, and increasing the data capacity in metropolitan regions, such spatial networks are expected to offer global Internet for all and, at the same time, provide terrestrial wireless backhaul solutions. Assessing the true benefits arising from integrating various single-layer networks at different altitudes (such as satellites, HAPs, and LAPs) remains, however, subject to several physical hurdles. Unlike terrestrial networks, high latency, constrained resources, mobility, and intermittent links are major spatial network issues, and so it becomes vital to study the interconnecting P2P links among various layers of spatial networks. To this end, this paper surveys the state-of-the-art on enabling P2P links in different layers of spatial networks. The paper first introduces spatial networks' background, including satellite, HAPs, and LAPs networks, and presents various exciting projects on the topic. Then, we explain two different solutions, i.e., RF and FSO, for connecting the satellites in a single orbit or at different orbits. We also present the link budget analysis for both RF and FSO-based satellite-to-satellite links.
Furthermore, we present the studies regarding RF and FSO for enabling HAP-to-HAP links and further explore the research on performance analysis of HAP networks. Afterward, we present the literature on integrated terrestrial and non-terrestrial networks as a means to  enable next-generation wireless communication systems. Finally, we identify numerous future research directions, including network optimization, intelligent offloading, smart platforms, energy efficiency, multiple access schemes, and distributed spatial networks. Up to the authors' knowledge, this is the first paper of its kind that surveys P2P links for a multi-layered spatial network in light of 6G large-scale complex networks. Many of the paper insights intend at enabling the establishment of P2P links in future integrated spatial networks.

\bibliographystyle{IEEEtran}
\bibliography{ref}

\end{document}